# Non-Markovian trajectories involving future in the semi-classical path integral expression

Fei Wang[1]

[1]Department of Chemistry, Le Moyne College, Syracuse, US

wangfe@lemoyne.edu

**Abstract**

Semi-classical path integral expression for a quantum system coupled to a harmonic bath is derived based on the stationary phase condition. It is discovered that the system path is non-Markovian. Most strikingly, the system path not only couples to its past, but also to its future, i.e. the equation of motion for the system is an integro-differential equation that involves all times. Numerical examples are given at the end. Because of the future-non-Markovian nature of the equation, the numerical solution cannot be obtained by iterative methods. Instead, root search algorithms must be employed.

Keywords: non-Markovian, semi-classical, path integral, quantum dynamics

## I. Introduction

The dynamics of a quantum system interacting with its environment plays a central role in many areas of research, from addressing fundamental questions such as quantum decoherence and quantum measurement[1-6], to many practical applications in vibrational energy relaxation in molecules[7-11], electron transport in condensed phase[12-15], excitonic energy transport in photosynthesis[16-20], spin-lattice relaxation[21, 22] and entanglement in quantum computing[23-25]. Due to the enormous degrees of freedom involved, a common strategy is to divide the entire system into a quantum system (a few degrees of freedom) and the bath[26]. For example, for electron transfer in proteins, the electron is considered to be the quantum system, and the proteins and the solvent are the bath.

Due the decoherence effect of the bath[27], the wavefunction is not a good representation of the system state. Instead, the quantum system is described by the density matrix[28, 29], i.e. the probability instead of the probability amplitude, because the phase will eventually be washed out by the bath so called dephasing or decoherence. Path integral method has proved to be a very useful scheme in describing the time evolution of the density matrix[27, 30]. Feynman's influence functional[26, 31, 32] of a quantum system coupled to a harmonic bath offers enormous insight into subsequent research investigations[27, 30, 33-38]. The reasons that harmonic bath has received most of the attention are its analytically tractable expressions[36, 39] and the Gaussian-fluctuation nature of the bath through collective modes[40]. One feature that the influence functional presents is the time non-local correlation function that renders non-Markovian dynamics.[36, 39, 41-49].

With the development of high-speed computers, more and more realistic condensed phase quantum systems have been simulated using Feynman's path integral scheme[50-56]. From a computer simulation point of view, to make Feynman's influence functional feasible, one key issue needs to be addressed: that is the infinite number of the quantum paths in the summation. For a two level system coupled to a harmonic bath, discrete variable representation (DVR)[57] and tensor propagator[58-60] have been proposed based on the small number of system states involved and finite time span of the correlation function. However, for a quantum system involving continuous variables (i.e. expectation value of a wave packet), semi-classical approaches are usually employed[61-63], in which the infinite number of paths is reduced onto the classical trajectories.

In this paper, I derive a semi-classical path integral expression for a quantum system coupled to a harmonic bath. The "semi-classical" specifically refers to the semi-classical trajectories in the quantum system coordinates. The following content is arranged as follows: In section II, I briefly introduce the path integra formulation. In section III, I concisely present the method for achieving semi-classical mechanics. In section IV, I derive the semi-classical path integral formulation of a system coupled to a harmonic bath, along with discussions of its properties. In section V, I present the numerical calculations. In section VI, I make conclusion remarks.

**II. Path integral**

Consider the Hamiltonian

$$H(s,p) = \frac{p^2}{2m} + V(s) \tag{2.1}$$

where $s$ represents position and $p$ represents momentum. Here the Hamiltonian is one dimensional for simplicity. The generalization to many degrees of freedom is straightforward.

Breaking up the time propagator into $N$ short time segments,

$$e^{-iHt/\hbar} = e^{-iH\Delta t/\hbar} e^{-iH\Delta t/\hbar} \ldots e^{-iH\Delta t/\hbar} \tag{2.2}$$

and inserting the complete set of position states,

$$\int ds_k |s_k\rangle\langle s_k| = 1 \tag{2.3}$$

the time propagator in this position representation becomes,

$$\langle s_f | e^{-\frac{iHt}{\hbar}} | s_0 \rangle = \int ds_1 \ldots \int ds_{N-1} \prod_{k=1}^{N} \langle s_k | e^{-\frac{iH\Delta t}{\hbar}} | s_{k-1} \rangle \tag{2.4}$$

where $s_0$ represents the initial position at time 0, and $s_f$ the position at time $t$.

Now employing the Trotter splitting[64]: the symmetric splitting of the potential energy term,

$$e^{-iH\Delta t/\hbar} \approx e^{-iV\Delta t/2\hbar} e^{-iT\Delta t/\hbar} e^{-iV\Delta t/2\hbar} \tag{2.5}$$

the time propagator becomes,

$$\langle s_k | e^{-iH\Delta t/\hbar} | s_{k-1} \rangle \approx \langle s_k | e^{-iT\Delta t/\hbar} | s_{k-1} \rangle e^{-(i\Delta t/2\hbar)[V(x_k)+V(x_{k-1})]} \tag{2.6}$$

The kinetic energy part can be evaluated analytically,

$$\langle s_k | e^{-iT\Delta t/\hbar} | s_{k-1} \rangle = \left( \frac{m}{2\pi i \hbar \Delta t} \right)^{1/2} \exp\left[ \frac{i}{\hbar} \frac{m}{2\Delta t} (s_k - s_{k-1})^2 \right] \tag{2.7}$$

Therefore,

$$\langle s_f | e^{-iHt/\hbar} | s_0 \rangle \approx \left( \frac{mN}{2\pi i \hbar t} \right)^{N/2} \int ds_1 \ldots \int ds_{N-1}$$
$$\exp\left\{ \frac{i}{\hbar} \left[ \frac{mN}{2t} \sum_{k=1}^{N} (s_k - s_{k-1})^2 - \frac{t}{N} \sum_{k=1}^{N} V(s_k) \right] \right\} \tag{2.8}$$

As $N \to \infty$,

$$\langle s_f | e^{iHt/\hbar} | s_0 \rangle = \int \mathcal{D} s_t \left( \exp \frac{i}{\hbar} \int_0^t L(s, \dot{s}) dt' \right) = \int \mathcal{D} s_t \left( \exp \frac{i}{\hbar} \varphi[s_t] \right) \tag{2.9}$$

where $\dot{s}$ represents the velocity, $L(s, \dot{s})$ is the Lagrangian and $\varphi[s_t]$ is the action integral. $\int \mathcal{D} s_t$ means summing (i.e. integrating over) all possible paths, hence the name *path integral*.

This brings to the celebrated form in which the classical action enters in the phase of the exponential. Different paths will give different values of action, thus different phase factors, and when adding them up, allowing interference that is characteristic of quantum mechanics. The stationary phase condition, $\delta\varphi[s_t] = 0$, gives Euler-Lagrange equations of motion, $\frac{\partial L}{\partial s} = \frac{d}{dt}\left(\frac{\partial L}{\partial \dot{s}}\right)$, which is the classical equation of motion. The beauty of the Feynman's path integral formulation is that it provides a natural connection between quantum and classical mechanics. It should be interesting and important to point out that the classical trajectory in the action integral for the harmonic oscillator gives the exact quantum result, with no approximations.

### III. Semiclassical propagator

Semiclassical mechanics[65] offers a unique advantage in treating quantum systems in that the infinite sum of trajectories in the path integral is reduced to a single classical trajectory. Even better than that, it preserves some degree of quantum effect. In other words, it puts quantum flesh on classical bones.

The essence of semiclassical theory is the stationary phase condition, i.e. expanding the action around classical trajectories in power series, and then truncating up to second order. Consider the path

$$s(t) = s_0(t) + \delta s(t) \tag{3.1}$$

where $s_0(t)$ corresponds to a classical trajectory. The action $\varphi$ can be expanded in orders of $\delta s(t)$ with fixed endpoints $(s',t';s,t)$

$$\varphi = \varphi_0 + \delta\varphi_0 + \frac{1}{2}\delta^2\varphi_0 + \ldots \tag{3.2}$$

where

$$\varphi_0 = \int_0^t L(s_0, \dot{s}_0, t') dt' \tag{3.3}$$

where $s_0$ and $\dot{s}_0$ are position and velocity following classical trajectories.

Since $\delta\varphi_0 = 0$, the semiclassical time evolution operator can be written as

$$\langle s'|e^{-iHt/\hbar}|s\rangle^{SC} = \left(\frac{m}{2\pi i\hbar t}\right)^{1/2} \sum_{\text{classial paths}} \int \mathcal{D}\delta s_0 e^{i\left(\varphi_0 + \frac{1}{2}\delta^2\varphi_0\right)/\hbar} \tag{3.4}$$

The second variation in the exponential is a Gaussian integral in position and can be integrated out, which brings to the form

$$\langle s'|e^{-iHt/\hbar}|s\rangle^{SC} = \sum_{\text{classial paths}} prefactor \times e^{i\varphi_0/\hbar} \tag{3.5}$$

The pre-factor will be stated more explicitly in the following sessions.

## IV. Semi-classical path integral for a system coupled to a harmonic bath

The Hamiltonian for a system coupled to a harmonic bath can be written as

$$H = \frac{P^2}{2M} + V_0(s) + \sum_{i=1}^{N} \left( \frac{p_i^2}{2m_i} + \frac{m_i \omega_i^2}{2} x_i^2 - c_i f_i(s) x_i + \frac{c_i^2 f_i^2(s)}{2 m_i \omega_i^2} \right) \tag{4.1}$$

The letter $s$ represents the system coordinate and $x_i$ represents the bath coordinate for each oscillator. $c_i$ is the coupling strength. The last term is the counter term added to counterbalance the distortion of the system potential by the interaction.

The bath degree of freedom behaves like a forced oscillator with the equation of motion

$$m\ddot{x} + m\omega^2 x = cf(s) \tag{4.2}$$

The solution of $x(t)$ can be obtained by the standard Green's function technique,

$$x(t) = x_0 \cos \omega t + \frac{p_0}{m\omega} \sin \omega t + \frac{c}{m\omega} \int_0^t dt' f(s(t')) \sin \omega(t - t') \tag{4.3}$$

This is the classical result. However, just including the classical (forced) harmonic oscillator expression in the path integral expression (instead of sum of infinite trajectories) gives exact quantum result. This is the beauty of the harmonic oscillator.

The time evolution of the entire system is described by the density matrix,

$$\begin{aligned}
\langle s^+, x^+ | \rho(t) | s^-, x^- \rangle &= \langle s^+, x^+ | e^{-i\hat{H}t/\hbar} \rho(0) e^{+i\hat{H}t/\hbar} | s^-, x^- \rangle \\
&= \int ds_0^+ dx_0^+ ds_0^- dx_0^- \langle s^+, x^+ | e^{-i\hat{H}t/\hbar} | s_0^+, x_0^+ \rangle \langle s_0^+, x_0^+ | \rho(0) | s_0^-, x_0^- \rangle \langle s_0^-, x_0^- | e^{+i\hat{H}t/\hbar} | s^-, x^- \rangle \\
&= \int ds_0^+ dx_0^+ ds_0^- dx_0^- \rho(s_0^\pm, x_0^\pm) \langle s^+, x^+ | e^{-i\hat{H}t/\hbar} | s_0^+, x_0^+ \rangle \langle s_0^-, x_0^- | e^{+i\hat{H}t/\hbar} | s^-, x^- \rangle \\
&= \int ds_0^+ dx_0^+ ds_0^- dx_0^- \rho(s_0^\pm, x_0^\pm) Q(t)
\end{aligned} \tag{4.4}$$

where $\rho(s_0^\pm, x_0^\pm) \equiv \langle s_0^+, x_0^+ | \rho(0) | s_0^-, x_0^- \rangle$ is the initial state of the system-bath composite system, and the particular expression is at our choice and will be specified in the numerical simulation session V. $Q(t) \equiv \langle s^+, x^+ | e^{-i\hat{H}t/\hbar} | s_0^+, x_0^+ \rangle \langle s_0^-, x_0^- | e^{+i\hat{H}t/\hbar} | s^-, x^- \rangle$ is the time propagator of the density matrix. Note that the time propagator $Q(t)$ involves two terms: $\langle s^+, x^+ | e^{-i\hat{H}t/\hbar} | s_0^+, x_0^+ \rangle$ and $\langle s_0^-, x_0^- | e^{+i\hat{H}t/\hbar} | s^-, x^- \rangle$. The $\langle s^+, x^+ | e^{-i\hat{H}t/\hbar} | s_0^+, x_0^+ \rangle$ term represents the forward propagation, with a negative sign in front of the Hamiltonian, and the positions with the superscript + notation. The $\langle s_0^-, x_0^- | e^{+i\hat{H}t/\hbar} | s^-, x^- \rangle$ term represents the backward propagation, indicated by the positive sign in front of the Hamiltonian, and the negative

superscript on the positions. The total phase, i.e. the combined action integral of the forward-backward propagation, $\langle s^+, x^+ | e^{-i\hat{H}t/\hbar} | s_0^+, x_0^+ \rangle \langle s_0^-, x_0^- | e^{+i\hat{H}t/\hbar} | s^-, x^- \rangle$, is

$$\Phi = \int_0^t dt' \left[ \frac{1}{2} M \dot{s}^+(t')^2 - V_0(s^+(t')) - \frac{1}{2} M \dot{s}^-(t')^2 + V_0(s^-(t')) \right]$$

$$+ \sum_i c_i x_{i,0} \int_0^t dt' \left[ f_i(s^+(t')) - f_i(s^-(t')) \right] \cos \omega_i t' + \sum_i \frac{c_i p_{i,0}}{m_i \omega_i} \int_0^t dt' \left[ f_i(s^+(t')) - f_i(s^-(t')) \right] \sin \omega_i t'$$

$$+ \sum_i \frac{c_i^2}{2 m_i \omega_i} \int_0^t dt' \int_0^{t'} dt'' \left[ f_i(s^+(t')) - f_i(s^-(t')) \right] \left[ f_i(s^+(t'')) + f_i(s^-(t'')) \right] \sin \omega_i (t' - t'')$$

$$- \sum_i \int_0^t dt' \frac{c_i^2}{2 m_i \omega_i^2} \left[ f_i^2(s^+(t')) - f_i^2(s^-(t')) \right] \quad (4.5)$$

If we assume separable initial distribution of the system and the bath, and the bath takes on Wigner distribution, then with (4.5) plugged into (4.4) and integrate over Wigner distribution, we recover Feynman-Vernon influence functional[34].

The stationary phase conditions are given by setting the derivative of $\Phi$ with respect to system coordinates to zero, while keeping the endpoints $s^+, s^-$ fixed. The resulting equations are the *equations of motion* (The detailed derivation is given in Appendix A).

The forward trajectory $s^+(t')$ is given by,

$$-M \ddot{s}^+(t') - V_0'(s^+(t')) + \sum_i c_i x_{i,0} f_i'(s^+(t')) \cos \omega_i t' + \sum_i \frac{c_i p_{i,0}}{m_i \omega_i} f_i'(s^+(t')) \sin \omega_i t'$$

$$+ \sum_i \frac{c_i^2}{2 m_i \omega_i} f_i'(s^+(t')) \int_0^{t'} dt'' f_i(s^+(t'')) \sin \omega_i (t' - t'')$$

$$- \sum_i \frac{c_i^2}{2 m_i \omega_i} f_i'(s^+(t')) \int_{t'}^{t} dt'' f_i(s^+(t'')) \sin \omega_i (t' - t'')$$

$$+ \sum_i \frac{c_i^2}{2 m_i \omega_i} f_i'(s^+(t')) \int_0^{t'} dt'' f_i(s^-(t'')) \sin \omega_i (t' - t'')$$

$$+ \sum_i \frac{c_i^2}{2 m_i \omega_i} f_i'(s^+(t')) \int_{t'}^{t} dt'' f_i(s^-(t'')) \sin \omega_i (t' - t'') - \sum_i \frac{c_i^2}{m_i \omega_i^2} f_i(s^+(t')) f_i'(s^+(t')) = 0 \quad (4.6)$$

where $f'$ means taking the first derivative of $f$, and $\ddot{s}$ means the second derivative with respect to time. $t'$ is the current time.

The backward propagation $s^-(t')$ will have the same expression as in (4.6) except $s^+$ replaced by $s^-$.

Note that the forward trajectory depends on the backward trajectory (i.e. the equation of motion for $s^+(t')$ involves $s^-(t'')$) and vice versa. In addition, these equations of motion not only couple the present time to the past (i.e. the integral $\int_0^{t'} dt''$), but also to the future time[66] (i.e. the integral $\int_{t'}^{t} dt''$). It is a surprising result at first glance that seems to violate causality. Careful inspection reveals that the stationary phase condition applies to all the intermediate system positions, but with the endpoints fixed. Therefore, the problem is formulated as a boundary value problem, such that the intermediate positions are predetermined by the endpoints. To solve these integro-differential equations, the iterative numerical methods for differential equations cannot be used. Instead, the root search algorithms must be employed.

The second variation of $\Phi$ leads to a Gaussian integral that produces a pre-factor. This pre-factor is cumbersome to write out but easy to implement into computer code in the discretized version. The final expression for the semiclassical density matrix propagator $Q^{SC}$ (replacing Q is (4.4)) is

$$Q^{SC}(t) = prefactor \times \exp\left\{\frac{i}{\hbar}\Phi_{cl}\left(x_0, p_0; s_c^{\pm}\right)\right\} \tag{4.7}$$

where $s_c^{\pm}$ stands for classical trajectories following equation (4.6), and the pre-factor is

$\left(\frac{M}{2\pi\hbar\Delta t}\right)^N \sqrt{\frac{(2\pi i)^{2N-2}}{\det\left(\frac{\delta^2\Phi}{\delta s_m^{\pm}\delta s_n^{\pm}}\right)}}$. The notation $\delta s_m^{\pm}\delta s_n^{\pm}$ denotes all possible combinations of forward and backward coordinate in the discrete manner. (See appendix B for detailed expressions)

## V. Numerical results

### A. Harmonic oscillator coupled to harmonic bath

The first step to test the future-involved term in the equation of motion is to compare the numerical result with analytical solutions. The simplest case is a harmonic oscillator bilinearly coupled to a harmonic bath. The Hamiltonian is

$$H = \frac{P^2}{2M} + \frac{1}{2}M\Omega^2 s^2 + \sum_{i=1}^{N}\frac{p_i^2}{2m_i} + \frac{m_i\omega_i^2}{2}x_i^2 - \sum_{i=1}^{N}c_i s x_i + \frac{c_i^2 s^2}{2m_i\omega_i^2} \tag{5.1}$$

The exact analytical result can be achieved through normal mode transformation of the system and bath coordinate combined, then propagating the free oscillators, and in the end transforming them back. An additional advantage concerning a collection of harmonic oscillators is that the semi-classical result is the same as the exact result, which, if the future-term is true, will perfectly match the exact result.

To proceed, the author calculated the expectation value of the position operator $\langle \hat{s} \rangle$ of the system initially prepared to have the ground state wave function of the harmonic oscillator, displaced away from the equilibrium position

$$\psi_0(s) = \left(\frac{M\Omega}{\pi\hbar}\right)^{\frac{1}{4}} e^{-M\Omega(s-a)^2/2\hbar} \tag{5.2}$$

The system starts with the density matrix

$$\langle s_0^+ | \rho_s(0) | s_0^- \rangle = \sqrt{\frac{M\Omega}{\pi\hbar}} e^{-M\Omega(s_0^+ - a)^2/2\hbar} e^{-M\Omega(s_0^- - a)^2/2\hbar} \tag{5.3}$$

The bath is initially at thermal equilibrium assuming Wigner distribution

$$\frac{1}{\hbar\pi} \tanh\left(\frac{1}{2}\hbar\omega\beta\right) \exp\left[-\tanh\left(\frac{1}{2}\hbar\omega\beta\right)\left(\frac{m\omega x^2}{\hbar} + \frac{p^2}{m\omega\hbar}\right)\right] \tag{5.4}$$

and the spectral density

$$J(\omega) = \frac{\pi}{2} \hbar \xi \omega e^{-\omega/\omega_c} \tag{5.5}$$

which mimics Debye solids that have a cut-off frequency.

Below is the result with system frequency $\Omega = 1$, mass $M = 1$, displacement $a = 1$, and bah parameters $\omega_c = 6$, $\xi = 2$, $\beta = 1$. The black curve is the accurate calculation and the dots are from the non-local differential equation using a root search algorithm. The matching of the two confirms that the stationary-phase differential equation does involve a future term.

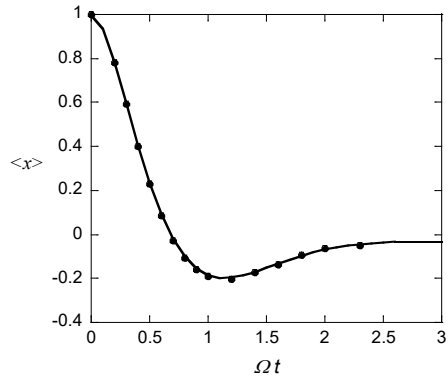

**B. Morse oscillator coupled to harmonic bath**

The author also extended the method to a non-harmonic quantum system, which desires to show whether the semi-classical method comprising the future-involved term serves as a good approximation. To compare results, a benchmark calculation from MCTDH (multi-configurational time-dependent Hartree) by H.D. Meyer[67] is utilized. The model is a Morse oscillator nonlinearly coupled to a harmonic bath.

$$H = \frac{P^2}{2M} + D\left(e^{-2as} - 2e^{-as}\right) - \sum_{i=1}^{N} c_i \frac{1-e^{-as}}{a} x_i + \sum_{i=1}^{N} \frac{p_i^2}{2m} + \frac{m\omega_k^2}{2} x_i^2 \quad (5.6)$$

where $s$ is the system coordinate and $x$ is the bath coordinate.

The harmonic frequency associated with the Morse potential is given by

$$\Omega = a\sqrt{\frac{2D}{M}} \quad (5.7)$$

Characteristic length scales of the Morse oscillator are given by $\bar{s} = 0.09129$, $D = 0.018$, $a = 2$, $M = 10^5$, $m = 10^4$. The system initial state is a displaced Gaussian wave packet with displacement $s_0 = 2\bar{s}$ and width $\sigma = \bar{s}$.

The spectral density uses Ohmic spectral density

$$J(\omega) = M\gamma\omega \quad (5.8)$$

with the following discretization, i.e. the modes are equally spaced.

$$\omega_i = i\frac{\omega_f}{N} = i\Delta\omega \quad (5.9)$$

$$c_i = i\sqrt{\frac{2mM\gamma\Delta\omega^3}{\pi}} \quad (5.10)$$

With $\gamma = \frac{1}{50\,fs}$. 20 oscillators are sufficient to converge the result.

The bath is at zero temperature and the oscillators are initially placed at the equilibrium position with respect to the system

$$\frac{\partial V}{\partial x_i} = 0 \quad (5.11)$$

$$x_i^{eq} = \frac{c_i}{m\omega_i^2} \frac{1-e^{-as}}{a} \quad (5.12)$$

The author used root search algorithm to perform the calculation of the expectation value of the system position. The initial guess positions are taken from the corresponding harmonic system linearly coupled to the harmonic bath. The black curve is the accurate calculation. The dots are from the semi-classical result, which matches the accurate result very well.

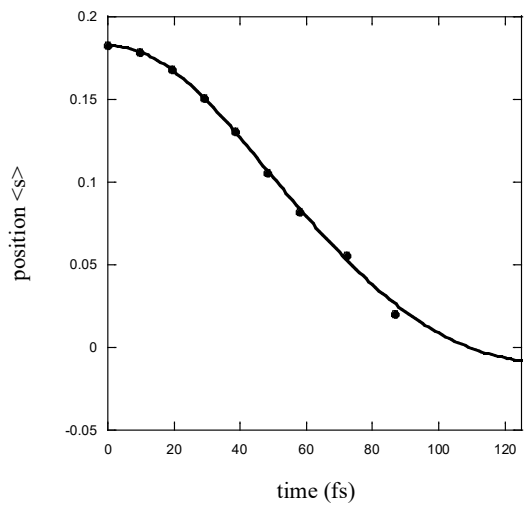

## VI. Conclusion

The semi-classical path integral expression for a quantum system coupled to a harmonic bath is derived. The quantum system trajectory is confined to a single classical trajectory, which greatly reduces the computational cost. Due to the formulation as a boundary value problem, the classical equation of motion for the system has a future-involved term. Root searching algorithms become essential for finding such system trajectories. Numerical examples are given the demonstrate that the semi-classical approximation can be a numerically sound method. It is worth mentioning that alternative to the boundary-value formulation is the initial value semi-classical formulation using coherent states, which is free from the future term. Active investigation is under way in my group.

**Acknowledgements**

The author thanks Professor Nancy Makri at the University of Illinois Urbana Champaign for useful discussions.

# Appendix A

The variation of $\Phi$ with respect to the system coordinate $s$ involves structure like this

$$\int_0^t dt' \int_0^{t'} dt'' f(s(t')) f(s(t'')) \sin\omega(t'-t'') \tag{A.1}$$

The variation of the above yields

$$\delta \int_0^t dt' \int_0^{t'} dt'' f(s(t')) f(s(t'')) \sin\omega(t'-t'')$$

$$= \int_0^t dt' \int_0^{t'} dt'' \delta f(s(t')) f(s(t'')) \sin\omega(t'-t'') + \int_0^t dt' \int_0^{t'} dt'' f(s(t')) \delta f(s(t'')) \sin\omega(t'-t'')$$

$$= \int_0^t dt' \delta s(t') f'(s(t')) \int_0^{t'} dt'' f(s(t'')) \sin\omega(t'-t'') + \int_0^t dt' \int_0^{t'} dt'' f(s(t')) \delta s(t'') f'(s(t'')) \sin\omega(t'-t'') \tag{A.2}$$

The second term can be manipulated into the variation of $\delta s(t')$ instead of $\delta s(t'')$ by switching the integration order of $dt'$ and $dt''$

$$\int_0^t dt' \int_0^{t'} dt'' f(s(t')) \delta s(t'') f'(s(t'')) \sin\omega(t'-t'')$$

$$= \int_0^t dt'' \int_{t''}^t dt' f(s(t')) \delta s(t'') f'(s(t'')) \sin\omega(t'-t'')$$

$$= \int_0^t dt'' \delta s(t'') f'(s(t'')) \int_{t''}^t dt' f(s(t')) \sin\omega(t'-t'') \tag{A.3}$$

Switch the label of $t'$ and $t''$, the above equation becomes

$$\int_0^t dt' \delta s(t') f'(s(t')) \int_{t'}^t dt'' f(s(t'')) \sin\omega(t''-t') = -\int_0^t dt' \delta s(t') f'(s(t')) \int_{t'}^t dt'' f(s(t'')) \sin\omega(t'-t'') \tag{A.4}$$

So, equation (A.2) becomes,

$$\delta \int_0^t dt' \int_0^{t'} dt'' f(s(t')) f(s(t'')) \sin\omega(t'-t'')$$

$$= \int_0^t dt' \delta s(t') f'(s(t')) \int_0^{t'} dt'' f(s(t'')) \sin\omega(t'-t'')$$

$$- \int_0^t dt' \delta s(t') f'(s(t')) \int_{t'}^t dt'' f(s(t'')) \sin\omega(t'-t'') \tag{A.5}$$

Therefore, the variation of the action can be written as (omitting the summation sign for simplicity),

$$\delta\Phi = \int_0^t dt' \delta s^+(t') \left[ -M\ddot{s}^+(t') - V_0'(s^+(t')) \right] - \int_0^t dt' \delta s^-(t') \left[ -M\ddot{s}^-(t') - V_0'(s^-(t')) \right]$$

$$+cx_0\int_0^t dt'\left[\delta s^+(t')f'(s^+(t'))-\delta s^-(t')f'(s^-(t'))\right]\cos\omega t'$$

$$+\frac{cp_0}{m\omega}\int_0^t dt'\left[\delta s^+(t')f'(s^+(t'))-\delta s^-(t')f'(s^-(t'))\right]\sin\omega t'$$

$$+\frac{c^2}{2m\omega}\int_0^t dt'\delta s^+(t')f'(s^+(t'))\int_0^{t'} dt''f(s^+(t''))\sin\omega(t'-t'')$$

$$-\frac{c^2}{2m\omega}\int_0^t dt'\delta s^+(t')f'(s^+(t'))\int_{t'}^t dt''f(s^+(t''))\sin\omega(t'-t'')$$

$$+\frac{c^2}{2m\omega}\int_0^t dt'\delta s^+(t')f'(s^+(t'))\int_0^{t'} dt''f(s^-(t''))\sin\omega(t'-t'')$$

$$-\frac{c^2}{2m\omega}\int_0^t dt'\delta s^-(t')f'(s^-(t'))\int_{t'}^t dt''f(s^+(t''))\sin\omega(t'-t'')$$

$$-\frac{c^2}{2m\omega}\int_0^t dt'\delta s^-(t')f'(s^-(t'))\int_0^{t'} dt''f(s^+(t''))\sin\omega(t'-t'')$$

$$+\frac{c^2}{2m\omega}\int_0^t dt'\delta s^+(t')f'(s^+(t'))\int_{t'}^t dt''f(s^-(t''))\sin\omega(t'-t'')$$

$$-\frac{c^2}{2m\omega}\int_0^t dt'\delta s^-(t')f'(s^-(t'))\int_0^{t'} dt''f(s^-(t''))\sin\omega(t'-t'')$$

$$+\frac{c^2}{2m\omega}\int_0^t dt'\delta s^-(t')f'(s^-(t'))\int_{t'}^t dt''f(s^-(t''))\sin\omega(t'-t'')$$

$$-\frac{c^2}{m\omega^2}\int_0^t dt'\left[\delta s^+(t')f(s^+(t'))f'(s^+(t'))-\delta s^-(t')f(s^-(t'))f'(s^-(t'))\right]$$

(A.6)

By setting the (A.6) to zero, we can immediately get the equations of motion (4.6).

**Appendix B**

To derive the second variation and integrate out the Gaussian integral, it is much easier to discretize the continuous system trajectories into smaller segments. Here we discretize the trajectory in the spirit of Trotter splitting, where we designate $s_0$ be the position from time 0 to $\frac{1}{2}\Delta t$, $s_1$ be the position from $\frac{1}{2}\Delta t$ to $\frac{3}{2}\Delta t$, ..., $s_k$ be the position for time $\frac{2k-1}{2}\Delta t$ to $\frac{2k+1}{2}\Delta t$, ..., $s_N$ be the position from $\frac{2N-1}{2}\Delta t$ to $N\Delta t$. The superscripts + and – denote forward and backward trajectories. Therefore, in discretized form, $\Phi = -\frac{1}{2}V_0\left(s_0^+\right)\Delta t - \sum_{k=1}^{N-1} V_0\left(s_k^+\right)\Delta t - \frac{1}{2}V_0\left(s_N^+\right)\Delta t + \frac{1}{2}V_0\left(s_0^-\right)\Delta t + \sum_{k=1}^{N-1} V_0\left(s_k^-\right)\Delta t + \frac{1}{2}V_0\left(s_N^-\right)\Delta t$

$+ \sum_{k=1}^{N} \frac{M}{2\Delta t}\left(s_k^+ - s_{k-1}^+\right)^2 - \sum_{k=1}^{N} \frac{M}{2\Delta t}\left(s_k^- - s_{k-1}^-\right)^2 + \frac{cx_0}{\omega}\left(f\left(s_0^+\right) - f\left(s_0^-\right)\right)\sin\left(\frac{1}{2}\omega\Delta t\right)$

$+ \frac{cx_0}{\omega}\sum_{k=1}^{N-1}\left(f\left(s_k^+\right) - f\left(s_k^-\right)\right)\left[\sin\omega\left(k+\frac{1}{2}\right)\Delta t - \sin\omega\left(k-\frac{1}{2}\right)\Delta t\right]$

$+ \frac{cx_0}{\omega}\left(f(s_N^+) - f\left(s_N^-\right)\right)\left[\sin\omega N\Delta t - \sin\omega\left(N-\frac{1}{2}\right)\Delta t\right]$

$+ \frac{cp_0}{m\omega^2}\left(f\left(s_0^+\right) - f\left(s_0^-\right)\right)\left(1 - \cos\frac{1}{2}\omega\Delta t\right)$

$+ \frac{cp_0}{m\omega^2}\sum_{k=1}^{N-1}\left(f\left(s_k^+\right) - f\left(s_k^-\right)\right)\left[\cos\omega\left(k-\frac{1}{2}\right)\Delta t - \cos\omega\left(k+\frac{1}{2}\right)\Delta t\right]$

$+ \frac{cp_0}{m\omega^2}\left(f\left(s_N^+\right) - f\left(s_N^-\right)\right)\left[\cos\omega\left(N-\frac{1}{2}\right)\Delta t - \cos\omega N\Delta t\right]$

$+ \frac{c^2}{2m\omega^3}\left(f\left(s_0^+\right) - f\left(s_0^-\right)\right)\left(f\left(s_0^+\right) + f\left(s_0^-\right)\right)\left(\frac{1}{2}\omega\Delta t - \sin\frac{1}{2}\omega\Delta t\right)$

$+ \frac{c^2}{2m\omega^3}\sum_{k=1}^{N-1}\left(f\left(s_k^+\right) - f\left(s_k^-\right)\right)\left(f\left(s_0^+\right) + f\left(s_0^-\right)\right)\left[\sin k\omega\Delta t - \sin(k-1)\omega\Delta t - \sin\left(k+\frac{1}{2}\right)\omega\Delta t + \sin\left(k-\frac{1}{2}\right)\omega\Delta t\right]$

$+ \frac{c^2}{2m\omega^3}\sum_{k=2}^{N-1}\sum_{k'=1}^{k-1}\left(f\left(s_k^+\right) - f\left(s_k^-\right)\right)\left(f\left(s_{k'}^+\right) + f(s_{k'}^-)\right)\left[2\sin(k-k')\omega\Delta t - \sin(k-k'-1)\omega\Delta t - \sin(k-k'+1)\omega\Delta t\right]$

$+ \frac{c^2}{2m\omega^3}\sum_{k=1}^{N-1}\left(f\left(s_k^+\right) - f\left(s_k^-\right)\right)\left(f\left(s_k^+\right) + f\left(s_k^-\right)\right)(\omega\Delta t - \sin\omega\Delta t)$

$+ \frac{c^2}{2m\omega^3}\left(f(s_N^+) - f\left(s_N^-\right)\right)\left(f\left(s_0^+\right) + f\left(s_0^-\right)\right)\left[2\sin\left(N-\frac{1}{2}\right)\omega\Delta t - \sin(N-1)\omega\Delta t - \sin N\omega\Delta t\right]$

$+ \frac{c^2}{2m\omega^2}\sum_{k=1}^{N-1}\left(f(s_N^+) - f\left(s_N^-\right)\right)\left(f\left(s_k^+\right) + f\left(s_k^-\right)\right)\left[\sin\left(N-k-\frac{1}{2}\right)\omega\Delta t - \sin(N-k-1)\omega\Delta t - \sin\left(N-k+\frac{1}{2}\right)\omega\Delta t\right.$
$\left. + \sin(N-k)\omega\Delta t\right]$

$+ \frac{c^2}{2m\omega^2}\left(f\left(s_N^+\right) - f\left(s_N^-\right)\right)\left(f\left(s_N^+\right) - f\left(s_N^-\right)\right)\left(\frac{1}{2}\omega\Delta t - \sin\frac{1}{2}\omega\Delta t\right)$

$- \frac{c^2}{2m\omega^2}\left(f^2\left(s_0^+\right) - f^2\left(s_0^-\right)\right)\frac{1}{2}\Delta t - \frac{c^2}{2m\omega^2}\sum_{k=1}^{N-1}\left(f^2\left(s_k^+\right) - f^2\left(s_k^-\right)\right)\Delta t - \frac{c^2}{2m\omega^2}\left(f^2\left(s_N^+\right) - f^2\left(s_N^-\right)\right)\frac{1}{2}\Delta t$   (B.1)

The second derivatives of $\Phi$ are given by the following

$$\frac{\partial^2 \Phi}{\partial s_k^{+2}} = -\Delta t V_0''(s_k^+) + \frac{2M}{\Delta t} + \frac{cx_0}{\omega} f''(s_k^+)\left[\sin\left(k+\frac{1}{2}\right)\omega\Delta t - \sin\left(k-\frac{1}{2}\right)\omega\Delta t\right]$$

$$+ \frac{cp_0}{m\omega^2} f''(s_k^+)\left[\cos\left(k-\frac{1}{2}\right)\omega\Delta t - \cos\left(k+\frac{1}{2}\right)\omega\Delta t\right]$$

$$+ \frac{c^2}{2m\omega^3} f''(s_k^+)(f(s_0^+) + f(s_0^-))\left[\sin k\omega\Delta t - \sin(k-1)\omega\Delta t - \sin\left(k+\frac{1}{2}\right)\omega\Delta t + \sin\omega\Delta t\right]$$

$$+ \frac{c^2}{2m\omega^3} f''(s_k^+) \sum_{k'=1}^{k-1}(f(s_{k'}^+) + f(s_{k'}^-))\left[2\sin(k-k')\omega\Delta t - \sin(k-k'-1)\omega\Delta t - \sin(k-k'+1)\omega\Delta t\right]$$

$$+ \frac{c^2}{m\omega^3} f(s_k^+) f''(s_k^+)(\omega\Delta t - \sin\omega\Delta t) + \frac{c^2}{m\omega^3} f'(s_k^+) f'(s_k^+)(\omega\Delta t - \sin\omega\Delta t)$$

$$+ \frac{c^2}{2m\omega^3} f''(s_k^+) \sum_{k'=k+1}^{N-1}(f(s_{k'}^+) - f(s_{k'}^-))\left[2\sin(k'-k)\omega\Delta t - \sin(k'-k-1)\omega\Delta t - \sin(k'-k+1)\omega\Delta t\right]$$

$$+ \frac{c^2}{2m\omega^3} f''(s_k^+)(f(s_N^+) - f(s_N^-))\left[\sin\left(N-k-\frac{1}{2}\right)\omega\Delta t - \sin(N-k-1)\omega\Delta t - \sin\left(N-k+\frac{1}{2}\right)\omega\Delta t + \sin(N-k)\omega\Delta t\right]$$

$$- \frac{c^2}{m\omega^2} f(s_k^+) f''(s_k^+) \Delta t - \frac{c^2}{m\omega^2} f'(s_k^+) f'(s_k^+) \Delta t, \qquad k = 1,\ldots, N-1 \qquad \text{(B.2)}$$

$$\frac{\partial^2 \Phi}{\partial s_k^+ \partial s_{k-1}^+} = -\frac{M}{\Delta t} + \frac{c^2}{2m\omega^3} f'(s_k^+) f'(s_{k-1}^+)(2\sin\omega\Delta t - \sin 2\omega\Delta t), \quad k = 2,\ldots, N-1 \qquad \text{(B.3)}$$

$$\frac{\partial^2 \Phi}{\partial s_k^+ \partial s_{k+1}^+} = -\frac{M}{\Delta t} + \frac{c^2}{2m\omega^3} f'(s_k^+) f'(s_{k+1}^+)(2\sin\omega\Delta t - \sin 2\omega\Delta t), \qquad k = 1,\ldots, N-2 \qquad \text{(B.4)}$$

$$\frac{\partial^2 \Phi}{\partial s_k^+ \partial s_{k'}^+} = \frac{c^2}{2m\omega^3} f'(s_k^+) f'(s_{k'}^+)\left[2\sin(k-k')\omega\Delta t - \sin(k-k'-1)\omega\Delta t - \sin(k-k'+1)\omega\Delta t\right],$$
$$1 \ll k' < k \ll N-1, \ k \neq k'+1 \qquad \text{(B.5)}$$

$$\frac{\partial^2 \Phi}{\partial s_k^+ \partial s_{k'}^+} = \frac{c^2}{2m\omega^3} f'(s_k^+) f'(s_{k'}^+)\left[2\sin(k'-k)\omega\Delta t - \sin(k'-k-1)\omega\Delta t - \sin(k'-k+1)\omega\Delta t\right],$$
$$1 \ll k < k' \ll N-1, \ k' \neq k+1 \qquad \text{(B.6)}$$

$$\frac{\partial^2 \Phi}{\partial s_k^+ \partial s_{k'}^-} = \frac{c^2}{2m\omega^3} f'(s_k^+) f'(s_{k'}^-)\left[2\sin(k-k')\omega\Delta t - \sin(k-k'-1)\omega\Delta t - \sin(k-k'+1)\omega\Delta t\right],$$
$$1 \ll k' < k \ll N-1 \qquad \text{(B.7)}$$

$$\frac{\partial^2 \Phi}{\partial s_k^+ \partial s_{k'}^-} = -\frac{c^2}{2m\omega^3} f'(s_k^+) f'(s_{k'}^-) \left[ 2\sin(k'-k)\omega \Delta t - \sin(k'-k-1)\omega \Delta t - \sin(k'-k+1)\omega \Delta t \right], \tag{B.8}$$

$$1 \ll k < k' \ll N-1$$

$$\frac{\partial^2 \Phi}{\partial s_k^+ \partial s_k^-} = 0, \quad k = 1, \ldots, N-1 \tag{B.9}$$

There will be exactly another 8 terms with $s_k^+$ and $s_k^-$ interchanged. Because they have exactly the same structure, I will not list them.

Each term of the above goes into the determinant symbolically written as $\det\left(\frac{\delta^2 \Phi}{\delta s_m^\pm \delta s_n^\pm}\right)$. The determinant of this $2N-2$ dimensional Jacobian can be evaluated numerically on a computer, e.g. through LU decomposition. This operation is $O(N^3)$. This Jacobian matrix is symmetric and can be divided into four equal-size square blocks. The upper left block is positive definite, and the lower right block is negative definite. They represent the coupling in forward-forward and backward-backward trajectories. The upper right and the bottom left blocks represent the couplings between the forward and backward trajectories. These couplings between positions (non-zero in the off-diagonal matrix elements) are manifestations of the non-Markovian dynamics involving both the past and future.